\documentclass{PoS}
\PoS{PoS(LAT2005)314}

\title{A toy model of (grand) unified monopoles}

\ShortTitle{A toy model of unified monopoles}

\author{\speaker{Lorenz von Smekal}\\

        Centre for the Subatomic Structure of Matter, School of
	Chemistry \& Physics,\\ The University of Adelaide,
	Adelaide, SA 5005, Australia  

        E-mail: \email{lorenz.smekal@adelaide.edu.au}}

\author{Torsten Tok\thanks{Supported by the {\em
        Deutsche Forschungsgemeinschaft} (DFG), contract
        SM 70/1-2.}\\

        Institut f\"ur Theoretische Physik, Universit\"at T\"ubingen,
        D-72076 T\"ubingen, Germany\\
        Institut f\"ur Theoretische Physik III, Universit\"at
        Erlangen, D-91058, Erlangen, Germany

        E-mail: \email{tok@tphys.physik.uni-tuebingen.de}}

\author{Philippe de Forcrand\\

        Institut f\"ur Theoretische Physik, ETH-Z\"urich,
        CH-8093 Z\"urich, Switzerland\\
	CERN Theory Division, CH-1211 Geneva 23, Switzerland

        E-mail: \email{forcrand@phys.ethz.ch}}

\abstract{We explore the old idea that, in a theory containing several
  gauge groups, the topological defects of one gauge group coincide
  with those of another gauge group. This simple 'unification'
  constraint has deep consequences, the best known of which is a
  natural explanation of the fractional electric charge of
  quarks. Here we explore the consequences of this idea for the phase
  diagram, in a toy model $U(1)\times U(1)$.}

\FullConference{XXIIIrd International Symposium on Lattice Field Theory\\

		 25-30 July 2005\\

		 Trinity College, Dublin, Ireland}

 \newcommand{\zr}[1]{\mbox{\hspace*{#1em}}}
 \newcommand{\ZZ}{\mbox{\sf Z\zr{-0.45}Z}}
 \newcommand{\slantfrac}[2]{\zr{.2}
             \raisebox{.5ex}{$\displaystyle
 #1$}\zr{-.3}/\raisebox{-.5ex}{$\displaystyle (#2)$}}

\begin{document}

\section{Introduction}

A while ago we measured in pure $SU(2)$ gauge theory the temperature 
behaviour of the free energy of various types of center vortices. 
Using 't~Hooft's twisted b.c.'s we studied ratios of partition functions 
with an odd number of center vortices piercing the various planes
of a 4-dimensional Euclidean $1/T\! \times\! L^3$ box relative to the
periodic ensemble. Qualitatively, at low temperatures, center vortices
can spread to lower their free energy. Their proliferation disorders
the Wilson loop and leads to confinement. As the temperature is
increased vortices through the temporal $1/T\! \times \! L $ planes 
are squeezed more and more. They can no-longer spread arbitrarily  and
this is what drives the phase transition. In the thermodynamic limit,
their free energy approaches zero (infinity) for $T$ below (above)
$T_c$ \cite{Kov00,Sme02,deF01}. In the high temperature phase, macroscopic 
regions of Polyakov loops of a definite center sector appear, which are
separated by interfaces whose tension suppresses these types of
vortices leading to a dual area law for the spatial 't~Hooft loops in
the high temperature phase \cite{deF00}. A Kramers--Wannier duality is observed
nicely in comparing the behaviour of these vortices with that of the
electric fluxes which yield free energies of static charges in a
well-defined (UV-regular) way~\cite{deF01b}, with boundary conditions
to mimic the presence of 'mirror' (anti)charges in neighbouring
boxes. This duality follows that between the Wilson loops of the 3d
$\ZZ_2$-gauge theory (the universal  partners of the spatial 't~Hooft
loops in $SU(2)$) and the 3d-Ising spins, reflecting the different
realisations of the 3-dimensional electric center symmetry in
both phases. Universality is seen at work in an impressively large
scaling window around criticality \cite{Sme02,Pep01}. 
While the electric (fluxes)twists provide well-defined (dis)order
parameters for confinement, the free energy of the magnetic ones
vanishes exponentially at all $T$ in the thermodynamic limit.
The corresponding screening of temporal 't~Hooft loops is determined
by the spatial string tension, and center monopoles always 'condense' 
\cite{Sme02b}. 
%There is no analogue of magnetic flux in the 3d-spin systems,  
%The corresponding 3d~{\em magnetic center symmetry} remains unbroken, 
Combinations of electric and magnetic twists can however be used to 
measure the topological susceptibility without cooling \cite{Sme02c}. 

While the techniques can be extended to compute interface tensions 
for $SU(N)$ with $N \ge 3$ and phase coexistence \cite{deF04},
the nagging question remains as to what the significance is of these
qualitatively and quantitatively quite compelling results
when dynamical quarks with their fundamental charges are included. 
Even in presence of quite heavy dynamical quarks the picture becomes rather
murky. Upon encircling an interface (which is a line-defect in 3
dimensions) they pick up a non-trivial phase corresponding to their fundamental
charge. This multivaluedness thus seems to have a dramatic effect on the
dynamics of the same topological defects that appear to describe 
the phases of the pure gauge theory so beautifully. Should it be true
that the phase structure changes abruptly when going from infinitely
heavy to no-matter-how-large but finite quark masses? At least it
would seem rather unnatural to assume that there are entirely different
mechanisms in either case, which nevertheless lead to a smooth limit.

In the next section we briefly discuss a tantalising possibility for
the coexistence of quarks and interfaces by simple unification
constraints in theories with several gauge groups as in
$SU(3)\times U(1)$. In Section 3 we explore this same idea in a toy
model $U(1) \times U(1)$. The considerable consequences of defect
unification as exemplified in the toy model are discussed in our
conclusions. The idea of unification constraints for topological
defects, which might arise quite naturally from grand unified theories,   
with topological defects such as e.g.~the $SU(5)$ monopole
\cite{Sre80}, is not new but certainly deserves renewed interest and
further study. 

\section{Quarks and Interfaces}

The problem with quarks and interfaces arises because a center-vortex
sheet, which can be moved freely in the pure gauge ensemble, now becomes
observable. Technically, with fundamental fields there are no twisted 
boundary conditions to measure vortex free energies in the first place.  
One way out of this dilemma might be provided by combining $SU(3)$ with  
$U(1)$ defects, realising that quarks have fractional electric charges
and interact with both gluons and photons. As pointed out by Creutz in
his last year's Lattice proceedings \cite{Cre04}, there is a remarkable
phase cancellation when quarks encircle a combined $\ZZ_3$ and Dirac
string. This is due to their fractional electric charges and makes
such combined strings unobservable.

In particular, upon encircling a $\ZZ_3$ string, quarks pick up phases
of $e^{i 2\pi/3}$ or $e^{i 4\pi/3}$ depending on the kind of string
(i.e. interface in 4d). On the other hand, depending on their
fractional electric charges, $2/3$ or $-1/3$ in units of $e$, these
two phases can add to produce multiples of $2\pi$ for the combined
string. In particular, independently of the quark's flavour this phase
cancellation happens for the first type of $\ZZ_3$ string whenever the
accompanying Dirac flux is $m= 1 \textrm{ mod } 3$ (in units of
$2\pi/e$), while for the second it requires  $ m = -1 \textrm{ mod }
3$ (one easily verifies that all other combinations lead to
non-integer fractions of $2\pi$ for the total phases of quarks with
either charge). 

Thus, if we combine in this way $SU(3)$ 't~Hooft with $U(1)$ monopole
loops, and the respective center-vortex/Dirac sheets
spanned by the two, the dynamics of the so combined defects might
allow to smoothly connect quark confinement to that of static
fundamental charges in the pure gauge theory. 
%This phase cancellation is not possible for all combinations.
%The {\em allowed} elementary ones are summarised in Table \ref{tab1}. 
%Since it would in principle be quite possible that a particular
%flavour does not see a particular type of unified monopole at all, one
%might even speculate that different kinds of combined defects confine 
%different quarks. 
Of course, the defect-unification constraint
induces a coupling between the two gauge groups. The dynamics of
the formation/suppression of one kind of defect is now tied  to the
other. We will explore this effect in a toy model.

\section{Toy Model}

\begin{figure}
\vspace*{-.4cm}
\hspace*{-.25cm}
\includegraphics[width=7.4cm]{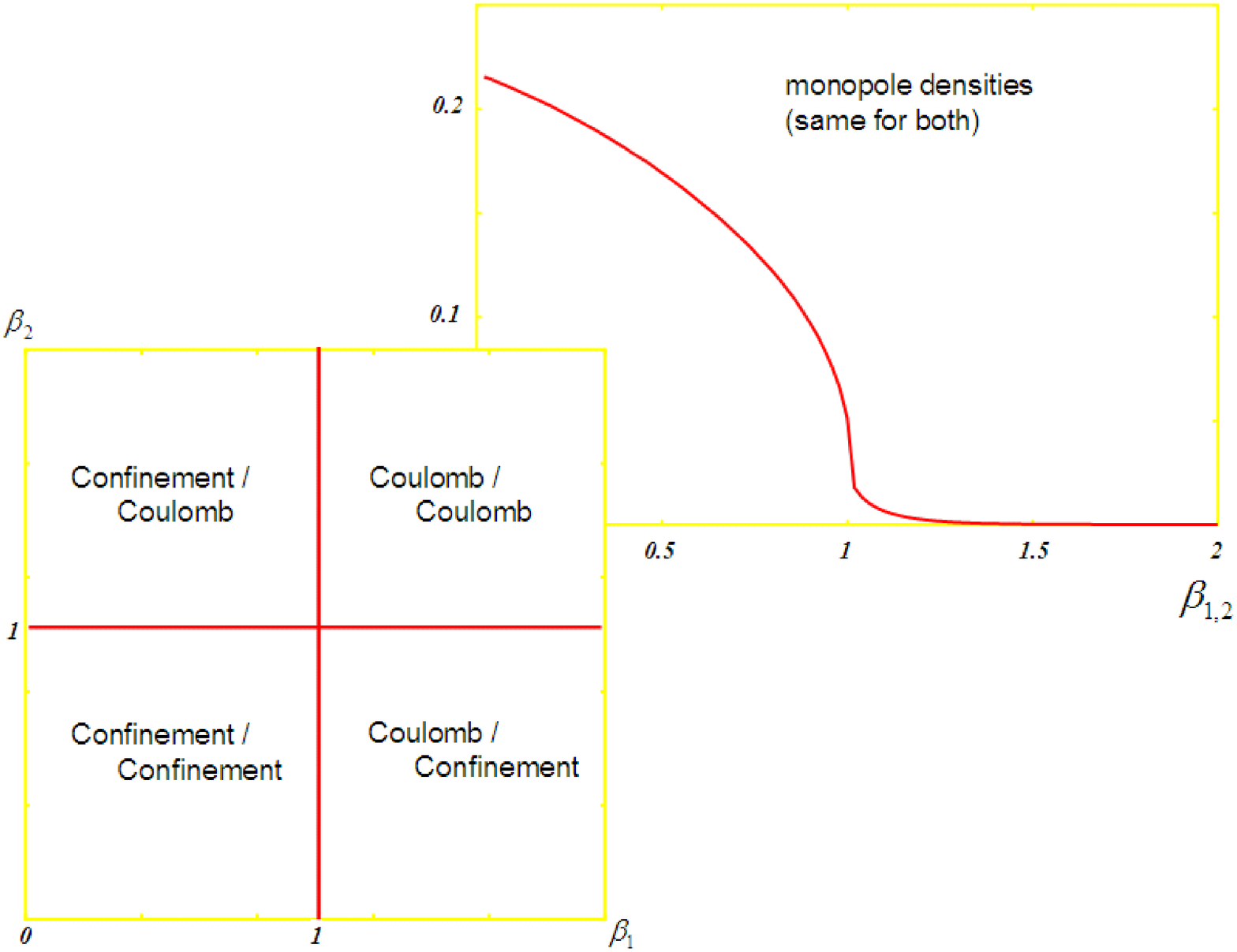}

\vspace{-5.6cm}
\leftline{\hspace{7.5cm}\includegraphics[width=8cm]{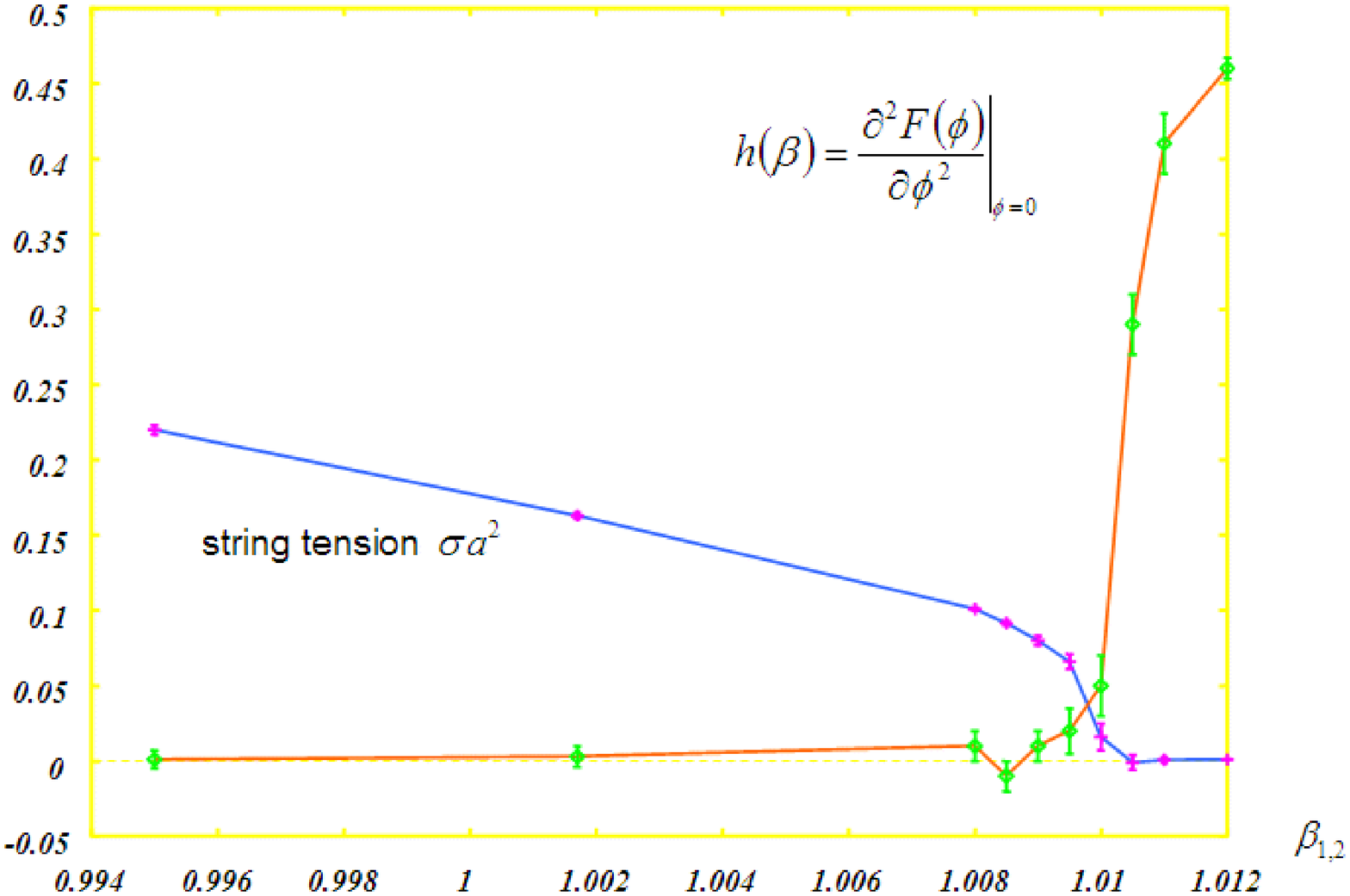}}
\vspace{.2cm}

\caption{Without constraint the monopole density (upper left), string
  tension and helicity modulus (right) of 4d compact $U(1)$ hold for
  each factor implying the 4 phases in the $(\beta_1,\beta_2)$-plane
  as sketched (left).}
\label{fig1}
\end{figure}

Consider the 4d compact pure Abelian gauge theory with Wilson action
for the gauge group $U(1)\times
U(1)$,
\begin{equation} 
            S \, =\, - \beta_1 \sum_P \cos \theta_P^{(1)} 
                  - \beta_2 \sum_P \cos \theta_P^{(2)} \; ,
\end{equation}
with two couplings $\beta_i = 1/e_i^2$ and plaquette
angles $\theta^{(i)}_P$, $i = 1,2$. Without any constraints the phases
are trivially determined by the 2 independent $U(1)$ factors. The
phase transitions at $\beta_c$ just above 1 can be seen in the
monopole densities, the string tension and the {\em helicity modulus}
\cite{Vet04} yielding the phase diagram as sketched in
Fig.~\ref{fig1}.  Especially the (temporal) 
helicity modulus has recently reemerged as a
suitable and convenient order parameter for compact $U(1)$
\cite{Vet04b}. It measures the susceptibility of the theory to static external
fluxes and plays a role analogous to that of fluxes by twisted boundary
conditions, albeit being more easily amenable to simulations.

The picture changes considerably when the two $U(1)$ factors are
constrained to always have the same monopole content. This is achieved
as follows: We start with identical copies of gauge field configurations 
for both $U(1)$'s. We then propose independent small variations for
each link of the two gauge fields a la Metropolis. In addition to the
usual probability by the Wilson action the monopole currents in both
$U(1)$'s must however also remain the same for the updates to be
accepted, though they may both be changed of course in each step (the
step size is chosen to achieve an acceptance rate of about 50\%).
This is controlled by comparing the units of magnetic
charge $m$ in the cubes sharing the updated links.

\begin{figure}
\hspace*{3.5cm}  
\parbox{11cm}{\hspace*{.4cm}\includegraphics[width=11.4cm]{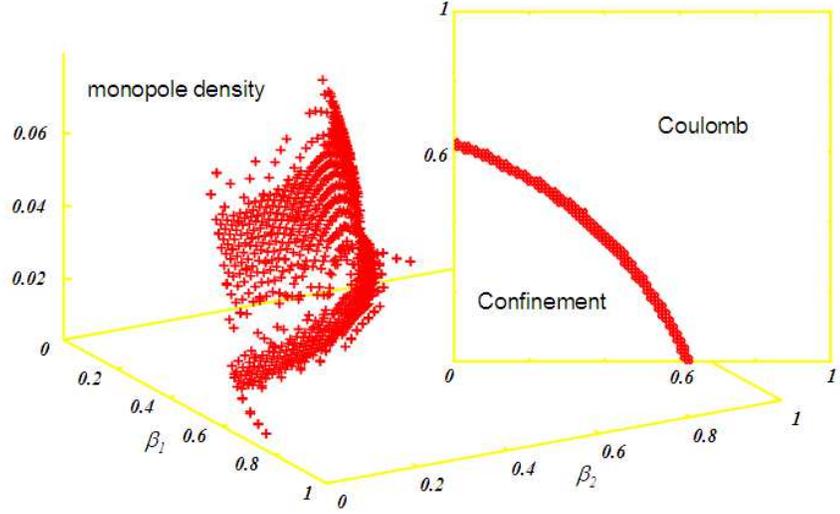}} 

\vspace{-7.25cm}
%\leftline{
\parbox{4cm}{\caption{Monopole density (relative number of cubes
    with $m\!\not=0$) and transition line estimate in the
    2 coupling plane of $U(1)\times U(1)$ with monopole unification
    constraint (on a $12^4$ lattice). 
    The relatively small probability of having a monopole in any given cube is
    further reduced in the constrained case, %.~This might explain why 
    and the transition to the confined phase is thus shifted towards
    the origin in the $(\beta_1,\beta_2)$-plane.}\label{fig2}} \hfill 
%}
\end{figure}

The phase diagram now changes dramatically as seen in
Fig.~\ref{fig2}.
One large enough $\beta$ (small coupling) suffices to suppress
the monopoles in both $U(1)$ groups, no matter how strong the second
coupling is.   
We are left with only 2 phases, the same for both $U(1)$'s, and the 
mixed phases no-longer exist. The transition occurs at somewhat
smaller values of $\beta$; we roughly estimate $0.39$ along the
diagonal $\beta_1 = \beta_2$ and $0.62$ near the axes where one of the
two $\beta$'s approaches zero. A precise determination remains to do be
done, however. Both string tensions (not shown here) 
are the same also for $\beta_1 \not= \beta_2$ and vanish at the
transition line in the two coupling plane. These correspond however to
Wilson loops of charges $(1,0)$ and $(0,1)$ (in units of $e$). 

\begin{figure}
\parbox{10cm}{\hspace*{-.2cm}\includegraphics[width=10cm]{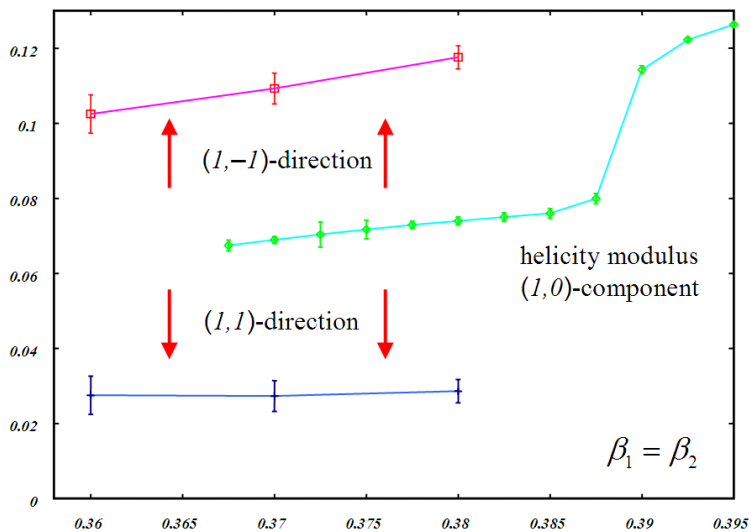}}
\hfill
\parbox{5cm}{\vspace*{-.3cm}
\caption{Susceptibility to flux in one of the two $U(1)$
factors %of the toy model 
with monopole-unification for $\beta_1=\beta_2$ 
%dircetion of the 2 coupling plane 
(on a $12^4$ lattice). 
Preliminary data for equal and opposite fluxes 
(the helicity moduli in $(1,\pm 1)$ directions) in the strong coupling
regime are also shown indicating a \emph{level splitting}.~A precise
analysis, a study of their transition behaviour 
and of finite-size effects are yet to be done. However, this is 
a first sign of the different behaviour of states with 
equal and opposite charges in both gauge groups
(which may each be half-odd integer).}\label{fig3}} 
\end{figure}

More interesting for our purposes are the half-odd integer cases
$(1/2,\pm1/2)$ which can exist in $U(1)\times U(1)$ with monopole
unification constraint because of the same phase cancellation as
described in Sec.~2. Instead of a direct measurement of the corresponding
Wilson loops, here we present preliminary results for the helicity moduli
which mimic the response to the presence of static charges. In
presence of constant (homogeneous) fluxes $\phi^{(i)}$ through
a $(\mu,\nu)$-plane (of size $L_\mu L_\nu$), the action of each $U(1)$
factor is modified for all plaquettes $P(\mu,\nu)$ with
$(\mu,\nu)$-orientation, 
\begin{equation}
S_i(\phi^{(i)}) = -\beta_i\zr{-1} \sum_{P\in\{P(\mu,\nu)\}} \zr{-1}  
\cos\big(\theta^{(i)}_P + \slantfrac{\phi^{(i)}}{L_\mu L_\nu} \big)
-\beta_i\zr{-1} \sum_{P\not\in\{P(\mu,\nu)\}} \zr{-1}  
\cos\theta^{(i)}_P \;\; , \quad i=1,\, 2 \; .
\end{equation}
Since the fluxes of both factors are independent, the helicity modulus
defined as the curvature of the free energy at vanishing flux now has
3 independent components according to the Hessian
\begin{equation}
  H(\beta_1,\beta_2) \, = \, \left(\frac{\partial^2
  F(\phi^{(1)},\phi^{(2)})}{\partial\phi^{(i)}\partial\phi^{(j)}}
  %\Big|_{\phi^{(1)}=\phi^{(2)}=0} \right)  \;. 
  \bigg|_{0} \right)  \;. 
\end{equation}
The diagonal components are the susceptibilities of the coupled theory
to flux in one of the two $U(1)$ factors alone. Their behaviour should
reflect that of the $(1,0)$ and $(0,1)$ Wilson loops. For objects
equally(oppositely) charged  w.r.t.~both, we need the curvature of
free energy in the $(1,\pm 1)$ directions corresponding to helicity
moduli $h^\pm = H_{11} + H_{22} \pm 2 H_{12}$,
\begin{eqnarray}
h^\pm(\beta_1,\beta_2) &=& \frac{1}{(L_\mu L_\nu)^2} 
\left\{ \sum_{i=1,2} \Big( 
\Big\langle \beta_i\zr{-1} \sum_{P\in\{P(\mu,\nu)\}} \zr{-1}
\cos\theta^{(i)}_P \Big\rangle  - 
\Big\langle \Big(\beta_i\zr{-1} \sum_{P\in\{P(\mu,\nu)\}}
\zr{-1}\sin\theta^{(i)}_P \Big)^2 \Big\rangle \Big) \right.  \\ 
&& \hskip 4cm \mp\,  2 
\left. \Big\langle \Big( \beta_1\zr{-1} \sum_{P\in\{P(\mu,\nu)\}} \zr{-1}
\sin\theta^{(1)}_P \Big)\Big(
\beta_2\zr{-1} \sum_{P\in\{P(\mu,\nu)\}} \zr{-1}
\sin\theta^{(2)}_P) \Big)\Big\rangle \right\}  \; . \nonumber
\end{eqnarray}
Near the transition line we observe a drop of about 50\% 
for the helicity modulus in $(1,0)$, $(0,1)$, i.e.~for flux in
just one of the $U(1)$'s. Contrary to the unconstrained case, however, 
it does not drop to zero on the strong coupling side as seen for
$\beta_1 =\beta_2 $ in Fig.~\ref{fig3}. Our preliminary data on
$h^\pm(\beta_1,\beta_2)$ indicates, however, that this is due to the
very different behaviour of equal versus opposite fluxes.  Presumably,
fields with equal electric charges in both $U(1)$'s are 
%It might imply that fields with equal electric charges in both $U(1)$'s are 
confined by the condensation of the unified monopoles,  while
oppositely charged ones are not. This would open further interesting
possibilities. One might ask for instance, what happens if we relax
the constraint to include oppositely charged $\pm m$ 
magnetic monopoles sitting on top of each other? Would that confine
the opposite electric charges together with the equal ones?

\section{Conclusions}

Our toy model serves to demonstrate that in theories with several
gauge groups it can be deceiving to study the individual factors
separately, especially when there are mechanisms by which 
the defects of one gauge group are forced to coincide with those of
another. Such unification constraints have been suggested to arise
naturally from grand unified theories and might manifest themselves in
the presence of fundamental fields such as quarks in QCD or other
particles with fractional charges in several gauge groups. It appears
to be worthwhile to explore the idea of defect unification in other
models such as a double Abelian Higgs model with fractional charges,
or revisit $SU(2)\times U(1)$ with fundamental Higgs fields in the
light of defect-unification constraints.

Another interesting property of the toy model $U(1)\times U(1)$ is
its capacity to confine equal charge doublets whereas oppositely charged
ones do not see the unified monopoles but only the topologically trivial
gauge field fluctuations, and thus retain a Coulomb-like behaviour at any
coupling. At least, our preliminary data on the corresponding helicity
moduli is fully consistent with what one expects for such a behaviour.
%is fully consistent with  to reflect this
%behaviour of the different types of charge. 
A more quantitative analysis is under way.

Finally, and maybe most importantly, however, one should probably 
address such questions as: 
What kind of constraints can we get from defects in grand unified
theories, and what do we need to confine quarks but not electrons?

\providecommand{\href}[2]{#2}

\end{document}